\documentclass[aps,twocolumn,showpacs,amsmath,amssymb]{revtex4} 
 
\usepackage{color} 
\usepackage{graphicx}% Include figure files 
\usepackage{dcolumn}% Align table columns on decimal point 
\usepackage{bm}% bold math 

\newcommand{\Npart}{N_{\text{part}}} 
\newcommand{\Tdec}{T_{\text{dec}}} 
 
\newcommand{\GeV}{\rm{~GeV}} 
\newcommand{\fm}{\rm{~fm}} 
\newcommand{\mb}{\rm{~mb}} 
 
\begin{document} 
 
\title{Event-by-event hydrodynamics and elliptic flow from fluctuating initial state}

\author{H. Holopainen$^{a,b}$} 
\email{hannu.l.holopainen@jyu.fi}
\author{H. Niemi$^c$}
\email{niemi@th.physik.uni-frankfurt.de}
\author{K.J. Eskola$^{a,b}$}
\email{kari.eskola@phys.jyu.fi}
 
\affiliation{
$^a$Department of Physics, 
P.O.Box 35, FIN-40014 University of Jyv\"askyl\"a, Finland\\ 
$^b$Helsinki Institute of Physics, 
P.O.Box 64, FIN-00014 University of Helsinki, Finland\\
$^c$Frankfurt Institute for Advanced Studies, Ruth-Moufang-Str. 1, D-60438 Frankfurt am Main, Germany} 
 
\date{3 January, 2011} 
 
\begin{abstract} 
We develop a framework for event-by-event ideal hydrodynamics to study the differential
elliptic flow  which is measured at different centralities in Au+Au collisions at
Relativistic Heavy Ion Collider (RHIC). Fluctuating initial energy density profiles,
which here are the event-by-event analogues of the eWN profiles, are created using
a Monte Carlo Glauber model.  Using the same event plane method for obtaining $v_2$
as in the data analysis, we can reproduce both the measured centrality dependence and
the $p_T$ shape of charged-particle elliptic flow up to $p_T\sim2$~GeV. We also consider
the relation of elliptic flow to the initial state eccentricity using different
reference planes, and discuss the correlation between the physical event plane and the
initial participant plane. Our results demonstrate that event-by-event hydrodynamics
with initial state fluctuations must be accounted for before a meaningful lower limit
for viscosity can be obtained from elliptic flow data. 

\end{abstract} 
 
 
\maketitle 

\section{Introduction}
\label{intro}

Azimuthal anisotropy of final state particles produced in ultrarelativistic heavy ion
collisions can be used to measure the collective behavior of the dense particle system
formed in such collisions \cite{Ollitrault:1992bk}. The strong azimuthal anisotropy,
which has been observed in the transverse momentum spectra of hadrons in Au+Au collisions
at the Relativistic Heavy Ion Collider (RHIC) of the Brookhaven National Laboratory, is
also a signature of the formation of strongly interacting partonic matter, the
Quark-Gluon Plasma (QGP).

Ideal hydrodynamics has been successful in predicting and explaining the measured
elliptic flow in Au+Au collisions at RHIC \cite{Hirano:2001eu,Hirano:2002ds,
Huovinen:2001cy,Huovinen:2005gy,Huovinen:2007xh,Kolb:2000sd,Kolb:2001qz,Nonaka:2006yn,
Schenke:2010nt,Teaney:2001av,Teaney:2000cw,Niemi:2008ta}. Currently, a lot of effort
is devoted for developing a description of the QCD-matter evolution in terms of
dissipative hydrodynamics. The recent results show that even a small viscosity can
considerably decrease the elliptic flow \cite{Romatschke:2007mq,Luzum:2008cw,Song:2007fn,
Song:2007ux,Song:2009rh,Dusling:2007gi}.

However, all these ideal and viscous hydrodynamic studies tend to underestimate the
elliptic flow in most central collisions. Generally, the explanation for the deficit
has been thought to be the initial state density fluctuations which have not been
accounted for. In addition to taking into account the density fluctuations themselves,
special care should be taken in computing the elliptic flow with respect to the same
reference plane as in the data analysis. 

The initial state fluctuations can be implemented e.g. via a Monte Carlo Glauber (MCG)
model which makes possible to study the fluctuations of the initial matter eccentricity.
Geometric fluctuations in the positions of nucleons have been shown to increase the
initial eccentricity, which is then suggested to translate into elliptic flow of final
state particles \cite{Alver:2008zza}. Furthermore, the reference plane plays a crucial
role: the eccentricity is larger if one calculates it using the participant plane
(determined by the transverse positions of the participant nucleons and the beam axis)
instead of the reaction plane (determined by the impact parameter and the beam axis).

Recently, in Ref.~\cite{Hirano:2009ah}, hydrodynamical calculations were performed using
averaged initial density profiles which were obtained from MCG calculations. Before
averaging over the profiles, the transverse coordinate axes were rotated in each event
so that the participant planes were on top of each other. In this manner it is possible
to get an averaged initial profile that takes into account the eccentricity fluctuations
in the initial state. For Au+Au collisions at RHIC, however, the effects of such plane
rotations on the integrated $v_2$ were small. 

While the above studies are steps to the right direction, it is obvious that without
doing event-by-event hydrodynamic simulations, it is impossible to know how closely the
computational participant plane corresponds to the physical event plane which is
determined from the observed final state hadron momenta. 

So far, genuine event-by-event models where hydrodynamics is run event by event using
fluctuating initial density profiles, have been presented in Refs.~\cite{Hama:2004rr,
Andrade:2006yh,Andrade:2008xh,Werner:2010aa,Petersen:2009vx,Petersen:2010md}.
Interestingly, a similar two-particle correlation ridge as observed in the experiments
\cite{:2009qa}, is seen to form into the rapidity--azimuth-angle --plane both in
NeXSpherio \cite{Andrade:2008xh} and more recently in Ref.~\cite{Werner:2010aa}. This
suggests that the puzzling ridge may well be another consequence of the fluctuations
in the initial state.

Also higher flow coefficients have been measured \cite{Adams:2003zg,Abelev:2007qg,
:2010ux} and recent studies \cite{Gombeaud:2009ye} show that the initial state density
fluctuations may play an important role in understanding the centrality dependence
of the ratio $v_4/(v_2)^2$. Triangular flow arising from event-by-event fluctuations
\cite{Alver:2010gr} is also one of the things that should be studied further with
event-by-event hydrodynamics.

In this paper, we introduce an event-by-event ideal hydrodynamics framework to study
the following $v_2$-related problems: With ideal hydrodynamics using averaged initial
states, {\em (i)} there is a $v_2$ deficit in central collisions, as discussed above;
{\em (ii)} the shape and centrality dependence of $v_2(p_T)$ are unsatisfactory in that
the $p_T$ slopes of $v_2$ easily become too steep  and elliptic flow increases too much
towards noncentral collisions;
{\em (iii)} elliptic flow is computed relative to the initial reaction plane or in the
best case to the participant plane \cite{Hirano:2009ah} but not relative to the event
plane, which is commonly used in the experiments;
{\em (iv)} one does not know how closely the event plane and the initial participant
plane correspond to each other. A concrete illustration of the problems
{\em (i)}--{\em (ii)} can be found in Fig.~7.5. of Ref.~\cite{Niemi:2008zz}, and also
in Fig.~5 of our previous elliptic flow study \cite{Niemi:2008ta}.

We will show how event-by-event ideal hydrodynamics, initiated with a fluctuating
initial density profile obtained from a MC Glauber model, and especially the
determination of $v_2$ with respect to the event plane, conveniently solves the problem
of the $v_2$ deficit in the most central Au+Au collisions at RHIC. Simultaneously, we
can significantly improve the agreement with the data for $v_2$ at all centrality
classes up to 30-40\% most central collisions in the typical applicability region of
hydrodynamics, $p_T<2$~GeV. This in turn has the very important implication that viscous
effects can in fact be allowed to be smaller than previously thought. Finally, we also
show the correspondence between the event and participant planes and study the relation
between the elliptic flow and initial eccentricity using different reference planes. 

The rest of the paper is constructed as follows:  First, in Sec.~II we introduce our
framework for event-by-event hydrodynamics. Details discussed there are our MCG model,
computation of the fluctuating initial energy density profiles, MC modeling of thermal
spectra of final state hadrons, and MC modeling of the resonance decays. We also try to
discuss the points where our modeling could be improved. Section~III is devoted for
defining the event plane and elliptic flow. Also eccentricity issues are discussed there.
Our results are presented in Sec.~IV and conclusions are given in Sec.~V.

\section{Event-by-event hydrodynamics framework}
\label{sec:framework}

\subsection{MC Glauber model and centrality classes}
\label{subsec:MCG}

We use here a MCG model to define the centrality classes and to form initial states
with fluctuating density profiles. First, we distribute the nucleons in the colliding
nuclei randomly using the standard, spherically symmetric, two-parameter Woods-Saxon
(WS) nuclear density profile as the probability distribution. Our WS parameters for the
gold nucleus are  $R_A \approx 6.37\fm$ for the radius and $d = 0.54 \fm$ for the
surface thickness. In the transverse $(x,y)$ plane the two nuclei are separated by an
impact parameter $b$ between the centers of mass of the nuclei, which is determined
by sampling the distribution $dN/db \propto b$ in the region
$0\le b \le b_{\text{max}} = 20 \fm > 2r_0$. The longitudinal $z$ coordinate is taken
into account when sampling the initial nucleon positions but in what follows it does
not play any role.

Nucleons $i$ and $j$ from different nuclei are then assumed to collide if their
transverse distance is small enough,
\begin{equation}
  (x_i - x_j)^2 + (y_i-y_j)^2 \le \frac{\sigma_{NN}}{\pi},
\end{equation}
where $\sigma_{NN}$ is the inelastic nucleon-nucleon cross section. We apply here 
$\sigma_{NN} = 42 \mb$ for Au+Au collisions at $\sqrt{s_{NN}} = 200 \GeV$.

We note that our simple MCG model fails to reproduce the correlations between the
nucleons, since we use the WS distribution for determining the nucleon positions
independently from each other. In \cite{Broniowski:2010jd} it was observed that a
realistic model, which accounts for nucleon correlations \cite{Alvioli:2009ab}, can be
well approximated using an exclusion radius which prevents nucleon overlap. Using such
radius, or giving a finite size for the nucleons \cite{Hirano:2009ah}, causes deviations
from the WS distribution which should then be compensated by tuning of the parameters in
the initially sampled WS distribution.

To keep our modeling as transparent as possible we, however, choose not to apply an
exclusion radius or a nucleon size in our MCG model since according to
Ref.~\cite{Hirano:2009ah} only a 10\% uncertainty in the initial eccentricity can be
expected, which is a much smaller effect than e.g. the overall uncertainties related
to the choice of the initial density profiles.  

\begin{figure}[t]
  \includegraphics[height=9.0cm]{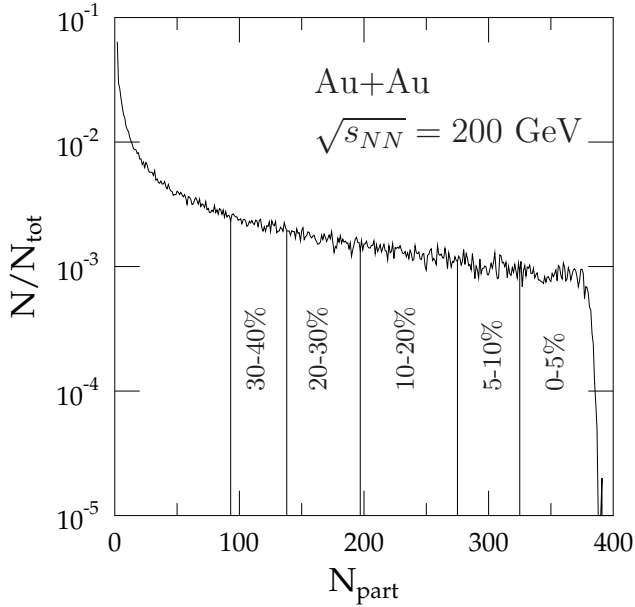}
  \caption{\protect\small Our definition of centrality classes for Au+Au
           collisions at $\sqrt{s_{NN}} = 200 \GeV$. Distribution
           of the number of participants is calculated from a Monte Carlo Glauber model
           without a nucleon exclusion radius.}
  \label{fig: centrality class}
\end{figure}

Next, we define the centrality classes using the number of participant nucleons,
$\Npart$, for simplicity. We have plotted the distribution of events as a function
of $\Npart$ in Fig.~\ref{fig: centrality class}. As indicated there, we slice our
total event distribution in $\Npart$ so that each $\Npart$ interval corresponds to
a centrality class which contains a certain percent of total events. The impact
parameter may thus freely fluctuate within each centrality class. 

\subsection{Initial density profiles}
\label{subsec:initial_densities}
In order to utilize the MCG-given initial state to start hydrodynamics, we must next
somehow transform the positions of the wounded nucleons or binary collisions into
energy density or entropy density. These would be the fluctuating event-by-event MCG
analogues of the conventional eWN, eBC and sWN, sBC \cite{Kolb:2001qz} average initial
densities. For simplicity, we consider here just the eWN-type of profile and leave the
profile fine tuning for future work. The energy density is now distributed in the
$(x,y)$ plane around the wounded nucleons using a 2D Gaussian as a smearing function,
\begin{equation}
  \epsilon (x,y) = \frac{K}{2\pi \sigma^2} \sum_{i=1}^{\Npart} \exp\Big( -\frac{(x-x_i)^2+(y-y_i)^2}{2\sigma^2} \Big),
  \label{eq:eps}
\end{equation}
where $K$ is a fixed overall normalization constant and $\sigma$ is a free smearing 
parameter controlling the width of our Gaussian. In each event, the impact parameter
defines the direction of the $x$ axis and the origin of the $(x,y)$ plane is determined
so that the energy-density weighted coordinate averages become
$\langle x \rangle = \langle y \rangle = 0\fm$.

For the hydrodynamical description to be meaningful, the initial state should not have
too sharp density peaks. In our MCG model we have given an effective interaction
radius $\sqrt{\sigma_{NN}/\pi}/2\approx 0.6\fm$ for the colliding nucleons, which sets
a natural order of magnitude for $\sigma$. To probe the sensitivity of our results to
the initial state smearing, we will consider two values, $\sigma=0.4\fm$ and 0.8~fm.
With the current setup, we cannot reduce $\sigma$ further, as this would require a
smaller step size in our hydrodynamical code, and consequently much more CPU time.
One should then also develop a way to handle multiple separate freeze-out surfaces,
see the discussion in Sec.~\ref{subsec:hydro}. These developments we leave as future
improvements. 

The reason to choose the energy density to be smeared rather than the entropy density,
is mostly technical and due to the fact that our focus here is on understanding the
transverse flow phenomena. Since we now avoid using the Equation of state in forming the
initial energy density profiles in each event, we have a more direct control on the
input energy density (pressure) gradients that drive the evolution of the transverse
flow and its asymmetries. In our case, the total energy per rapidity unit in each event,
$\int dxdy\,\epsilon(x,y)$, thus remains independent of $\sigma$, while the total
entropy per rapidity unit and thereby also the final state multiplicity depend on
$\sigma$. 

For Au+Au collisions at $\sqrt{s_{NN}} = 200 \GeV$, we use the value $K=37.8 \GeV/\fm$.
With this, we reproduce the initial total entropy of Ref.~\cite{Niemi:2008ta} when
averaging over many initial states in central ($b=0$) collisions when $\sigma=0.4\fm$.
Motivated by the EKRT minijet (final state) saturation model \cite{Eskola:1999fc} and
Ref.~\cite{Niemi:2008ta}, we fix the initial time to $\tau_0 = 0.17 \fm$ for all events.

\subsection{Hydrodynamics, freeze out and resonance decays}
\label{subsec:hydro}

For obtaining the ideal-fluid hydrodynamic evolution of the system, we solve the
standard equations
\begin{equation}
  \partial_\mu T^{\mu\nu} = 0
\end{equation}
together with an Equation of State (EoS) which relates pressure with the energy density
and net-baryon number density, $P=P(\epsilon, n_B)$. As we are interested in particle
production at mid-rapidity, we assume the net-baryon density to be negligible. Since
the rapidity distributions of hadrons are approximately flat at mid-rapidities we can
safely simplify our hydrodynamical equations by assuming longitudinal boost-invariance.
We solve this (2+1)-dimensional numerical problem using the SHASTA algorithm
\cite{Boris,Zalesak} which is also able to handle shock waves.

As the Equation of State (EoS), we choose the EoS from Laine and Schr\"oder
\cite{Laine:2006cp}. At high temperatures this EoS has been matched with the lattice-QCD
data and at low temperatures with a hadron resonance gas containing particles of mass
$m < 2 \GeV$. This EoS has a ''cross-over'' transition from the QGP to the hadron gas.

Thermal spectra for hadrons are calculated using the conventional Cooper-Frye method
\cite{Cooper}, where particle emission from a constant-temperature surface $\sigma$ is
calculated according to
\begin{equation}
  \frac{dN}{d^2 p_T dy} = \int_\sigma f(x,p) p^\mu d\sigma_\mu,
\label{eq:thermals}
\end{equation}
where $f(x,p)$ is the particle number-distribution function in momentum at a certain
space-time location. The freeze-out temperature $\Tdec=160$~MeV is fixed so that we
reproduce the measured $p_T$ spectrum of pions \cite{Adler:2003cb} when averaged initial
states are considered. 

Our surface finding algorithm operates in the $(r,\tau)$-plane for all spatial azimuthal
angles. Currently, we can find only surfaces which go through $r=0$. Due to the  initial
state fluctuations there might simultaneously exist also other, disconnected, freeze-out
surfaces which our algorithm does not recognize.  We have checked that for the
centrality classes and smearing parameters $\sigma$ considered here, only a a few
percent of the events actually contain such a surface. In any case, since these
additional surfaces typically originate from a few-nucleon collisions, they contribute
negligibly to particle production in not too peripheral Au+Au collisions. Making $\sigma$
smaller can also increase the number of disconnected freeze-out surfaces. To ensure the
applicability of our framework, we prefer not to consider centrality classes more
peripheral than 30--40\% or $\sigma<0.4\fm$ in the present study.

For the flow analysis, we need individual final state particles. In generating these
using the computed thermal spectrum as the probability distribution, we assume the total
number of thermal particles in a rapidity unit to be fixed individually in each event.
The transverse momentum $(p_x,p_y)$ for each particle is thus sampled from the
distribution $dN/d^2p_Tdy$ calculated in Eq.~(\ref{eq:thermals}).  Due to the assumed
boost-symmetry, we are not equipped to consider rapidity distributions, thus $y$ is
sampled from a flat distribution in the interval $|y| \le 0.5$.

Note that above we have neglected the fluctuations in the number of emitted thermal
particles. In principle one could derive these fluctuations separately from the thermal
distributions for each freeze-out surface element. However, it is not so clear how to
treat the space-like parts of the surface in this case. Since in the collisions
considered here there are of the order of 1000 particles per unit rapidity, these
fluctuations can in any case be expected to be negligible in comparison with the initial
state fluctuations.

Once we have generated all the thermal hadrons, we still need take into account the
strong and electromagnetic decays. We let the thermal resonances decay one by one
using PYTHIA 6.4 \cite{Sjostrand:2006za}. Some decay products can fall outside our
rapidity interval $|y|<0.5$. On the other hand, there would also be decay products
arriving from $|y|>0.5$ which we do not consider here. We have checked that instead of
increasing the width of our thermal particle rapidity window, to speed up the analysis,
we can simply count all decay products into our rapidity acceptance regardless of their
actual rapidity.

\subsection{Event statistics}
\label{subsec:statistics}

Our main goal is to compare the event-by-event hydrodynamic results with the ones
obtained by more conventional non-fluctuating hydrodynamics initiated with averaged
initial states. 

For event-by event hydrodynamics, we make 500 hydro runs in each centrality class. 
This amount of hydro runs seems enough for the hadron spectra and elliptic flow analysis. 
To increase statistics we make 20 final state events from every hydro run, thus we have
10 000 events in total. To check that using each hydro run 20 times is sensible, we have
checked that doing 250 hydro runs and 40 events from each leads to the same flow results. 

To create an averaged initial state, we sum together 20 000 initial states generated
by our MCG model. Such large number of events is required since fluctuations near the
edges of the system easily affect the final value of elliptic flow if the density
profile is otherwise smooth. We then do one hydro run with the averaged initial state
for each centrality class. To make a fare comparison with the event-by-event hydro
results, we do the resonance decays and analysis using the same code for the averaged
initial state case as for the event-by-event hydro case, making 10 000 final state
events from this one hydro run.

\section{Elliptic flow analysis}
\label{sec:v2}

\subsection{Elliptic flow and event plane}
\label{subsec:v2_event_plane}

The transverse momentum spectra of hadrons can be written as a Fourier series,
\begin{equation}
 \frac{dN}{d^2 p_T dy} = \frac{1}{\pi} \frac{dN}{dp_T^2 dy} \Big( 1 + 2\sum_{n=1}^\infty v_n \cos (n\phi) \Big),
\end{equation}
where $\phi$ is the hadron momentum's azimuthal angle with respect to the 
reaction plane defined by the impact parameter. The flow coefficients $v_n$ can then
be computed from
\begin{equation}\label{eq: v_n}
  v_n(p_T) = \frac{ \int d\phi \cos(n\phi) \frac{dN(b)}{dp_T^2 d\phi dy} }{ \int d\phi \frac{dN(b)}{dp_T^2 d\phi dy} }.
\end{equation}

When we have fluctuations in the initial state, calculation of $v_n$ is not so
straightforward. In the hydrodynamic runs, where we always know the direction of our
impact parameter, we can calculate the elliptic flow with respect to the reaction plane.
If we want to compare with experiments, we should use the same analysis methods and
definitions as in the data analysis. In this work we use the event plane method 
\cite{Poskanzer:1998yz,Ollitrault:1997di} which is a common way to calculate $v_2$. 
Since it is not (yet) typically used in hydrodynamical calculations, let us briefly
recapitulate the main points (see Ref.~\cite{Poskanzer:1998yz} for details).

We first define an event flow vector $\bm{Q_n}$ for the  $n$th harmonic. The event
flow vector in the transverse plane is
\begin{equation}
  \bm{Q_n} = \sum_i ( p_{Ti} \cos(n\phi_i),  p_{Ti} \sin(n\phi_i) ),
\end{equation}
where we sum over every particle in the event and where $\phi$ is measured from the
$x$ axis which is here fixed by the impact parameter. The event plane angle $\psi_{n}$
for each event is then defined to be 
\begin{equation}
  \psi_n = \frac{ \arctan ( Q_{n,y} / Q_{n,x} ) }{n},
\end{equation} 
with arctan placed into the correct quadrant. The "observed"  $v_n$ is calculated with
respect to the event planes obtained above,
\begin{equation}\label{eq: v2 obs}
  v_n\{\text{obs}\} = \langle \langle \cos (n(\phi_i - \psi_{n})) \rangle \rangle_{\text{events}},
\end{equation}
where the inner angle brackets denote an average over all particles $i$ in one event
and the outer ones an average over all events. In order to remove autocorrelations,
the particle $i$ is excluded from the determination of the event flow vector when
correlating it with the event plane. 

Since in our finite rapidity interval we have only a finite number of particles
available for the event plane determination, the obtained event plane fluctuates from
the "true" event plane. (In our event-by-event hydrodynamics, the true event plane in
each event would correspond to the average event plane obtained by generating infinitely
many final states from one hydro run.) The obtained $v_n\{\text{obs}\}$ is corrected
using the event plane resolution for the harmonic $n$ 
\begin{equation}
  \mathcal{R}_n = \langle \cos (n(\psi_n - \psi_n^{\text{true}})) \rangle,
\end{equation}
where $\psi_n^{\text{true}}$ defines the true event plane and the angle brackets stand
for an average over a large sample of events. Because experimentally it is not possible
to find the true event plane, the event plane resolution must be estimated. 

In the two-subevents method, which also we will use, each event is randomly divided
into two equal subevents $A$ and $B$. The event plane resolution for each of these
subevents is then  \cite{Poskanzer:1998yz} 
\begin{equation}
  \mathcal{R}_n^{\text{sub}} = \sqrt{ \langle \cos (n(\psi_n^A - \psi_n^B))
  \rangle}.
\end{equation}
If the fluctuations from the true event plane are Gaussian, one can analytically
obtain the following result \cite{Poskanzer:1998yz}
\begin{equation}\label{eq: ep resolution chi}
  \mathcal{R}_{n} 
  = \frac{\sqrt{\pi}}{2\sqrt{2}} \chi_n \exp (-\chi_n^2/4) \Big[ I_{0}(\chi_n^2/4) + I_{1}(\chi_n^2/4) \Big],
\end{equation}
where $I_0$ and $I_1$ are modified Bessel functions and $\chi_n \sim \sqrt{N}$, with
$N$ referring to the number of particles. Since we can calculate
$\mathcal{R}_n^{\text{sub}}$ from the subevents, we can numerically solve
$\chi_n^{\text{sub}}$ from Eq.~(\ref{eq: ep resolution chi}). Because the number of
particles in the subevents is half of those in the full events,
$\chi_n^{\text{full}} = \sqrt{2}\chi_n^{\text{sub}}$, and we can calculate the
resolution $\mathcal{R}_n^{\text{full}}$ for the full events. Finally, the flow
coefficients are obtained as
\begin{equation}
  v_n = \frac{v_n\{\text{obs}\}}{\mathcal{R}_n^{\text{full}}}. 
\end{equation}

The elliptic flow results computed with this method are denoted here as
$v_2\{\text{EP}\}$. We also compute the elliptic flow from Eq.~\eqref{eq: v2 obs} with
respect to the reaction plane using both fluctuating and averaged initial states. In
the reaction plane case we have no corrections coming from statistical fluctuations.
These results are denoted as $v_2\{\text{RP}\}$ in what follows. 

\subsection{Initial eccentricity and participant plane}
\label{subsec:eccentricity}

The reaction plane eccentricity of the hydrodynamical initial state can be defined as
(see e.g. Ref.~\cite{Alver:2008zza})
\begin{equation}
  \epsilon_{\text{RP}} = \frac{ \sigma_y^2 - \sigma_x^2}{\sigma_y^2 + \sigma_x^2}
\end{equation}
where
\begin{equation}
\begin{split}
  \sigma_y^2 &= \langle y^2 \rangle - \langle y \rangle^2 \\
  \sigma_x^2 &= \langle x^2 \rangle - \langle x \rangle^2, \\
\label{eq:eccRP}
\end{split}
\end{equation}
where the averaging is done over the energy density profile of Eq.~(\ref{eq:eps}).

Since the positions of wounded nucleons, however, fluctuate from one event to another,
tilting the transverse coordinate axes suitably we can actually get a larger
eccentricity than $\epsilon_{\text{RP}}$ above. Thus it is not so clear what the most
correct reference plane should be.

\begin{figure}[b]
  \includegraphics[height=9.0cm]{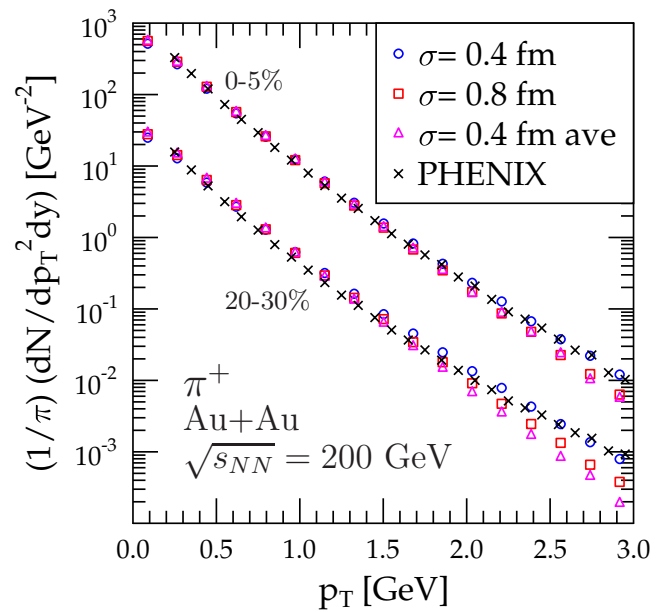}
\vspace{-1.0cm}
  \caption{\protect\small (Color online) Transverse momentum spectra of positive pions
           for Au+Au collisions at $\sqrt{s_{NN}} = 200 \GeV$ calculated with averaged
           and fluctuating initial states varying the width of Gaussian smearing.
           Data are from the PHENIX collaboration
           \cite{Adler:2003cb}.}
  \label{fig: pion spectra}
  \vspace{-1.0cm}
\end{figure}

\begin{figure*}[t!]
  \includegraphics[height=14.0cm]{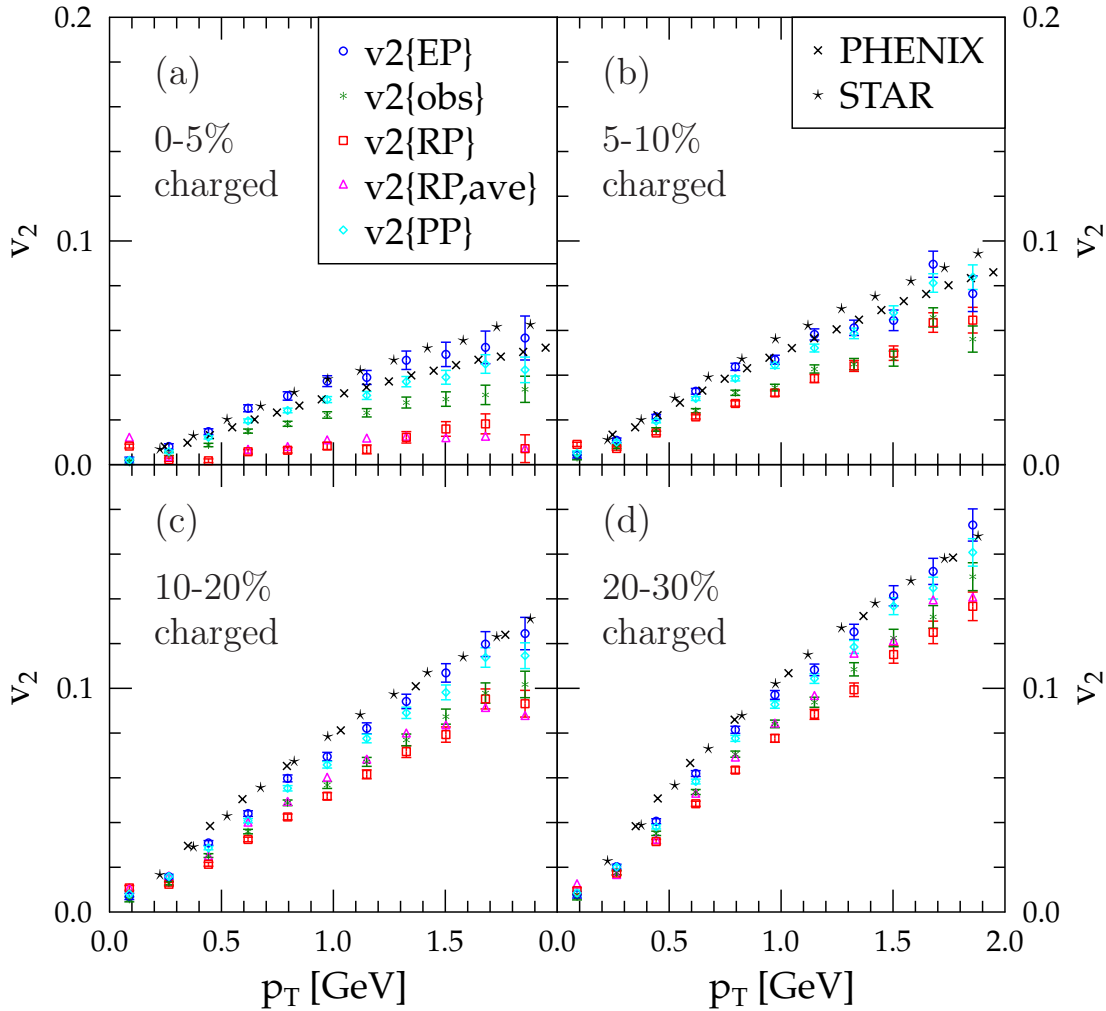}
  \caption{\protect\small (Color online) Elliptic flow of charged particles as a function
           of $p_T$ at different centralities for Au+Au collisions at $\sqrt{s_{NN}} =
           200 \GeV$. Hydrodynamical calculations with fluctuating and averaged
           initial states are shown for $\sigma = 0.4 \fm$. Data are
           from the PHENIX \cite{Afanasiev:2009wq,:2010ux} and STAR
           \cite{Adams:2004bi} collaborations. The statistical errors in the experimental
           data are smaller than the symbol size.}
  \label{fig: differential v2}
\end{figure*}

The reference plane that maximizes the initial eccentricity can be expected to correlate
better with the event plane than the reaction plane. For this purpose, one may define
the participant eccentricity \cite{Alver:2008zza}
\begin{equation}
  \epsilon_{\text{PP}} = \frac{\sqrt{(\sigma_y^2 - \sigma_x^2)^2 + 4\sigma_{xy}^2}}{\sigma_y^2 + \sigma_x^2},
\label{eq:eccPP}
\end{equation}
where $\sigma_{xy} = \langle xy \rangle - \langle x \rangle \langle y \rangle$. In this
case the reference plane is the participant plane which is defined by the $z$ axis
(beam direction) and the $x$ axis which is first rotated around the $z$ axis by the
angle
\begin{equation}
  \psi_{\text{PP}} = \arctan \frac{ - 2\sigma_{xy} }{ \sigma_y^2 - \sigma_x^2 +
              \sqrt{ (\sigma_y^2 - \sigma_x^2)^2 + 4 \sigma_{xy}^2 }  }.
\end{equation}

In what follows, we will compute the elliptic flow also with respect to the participant
plane,
\begin{equation}
v_2\{\text{PP}\}= \langle \langle \cos (2(\phi_i - \psi_{\text{PP}})) \rangle \rangle_{\text{events}}
\end{equation}
and consider the relation of elliptic flow to the initial eccentricity using both the
reaction plane and the participant plane as the reference.

\section{Results}

Below, we present the results for pion spectra, elliptic flow, eccentricities and the
correlation of the event and participant planes. The genuine event-by-event calculations
using smearings $\sigma=0.4$ and 0.8~fm, are compared with the results obtained using
an averaged initial state.

In Fig.~\ref{fig: pion spectra} we show the $p_T$ spectra of positive pions from these
three different hydro calculations and from the PHENIX collaboration \cite{Adler:2003cb}.
As explained in Sec.~\ref{subsec:initial_densities}, our multiplicity depends on the
Gaussian smearing width $\sigma$, hence the (small) difference between the points with
$\sigma = 0.4 \fm$ and $\sigma = 0.8 \fm$ at low $p_T$. 

We can also see that at higher $p_T$ we get more particles with the fluctuating
initial states than with the averaged initial state case. This follows from the fact
that in the fluctuating initial states there are larger pressure gradients present.
For the same reason, the high-$p_T$ spectra are quite sensitive to the value of $\sigma$:
with a larger $\sigma$, the pressure gradients are smaller and the $p_T$ spectra steeper.
This is in fact an interesting observation, suggesting that with fluctuating initial
states the applicability region of (event-by-event) hydrodynamics may extend to higher
$p_T$ than previously (see e.g. Refs. \cite{Eskola:2002wx,Eskola:2005ue}) thought.
In any case, the obtained $p_T$ spectra agree with the data sufficiently well, so that
we can meaningfully next study the elliptic flow.

\begin{figure}[b]
  \includegraphics[height=14.0cm]{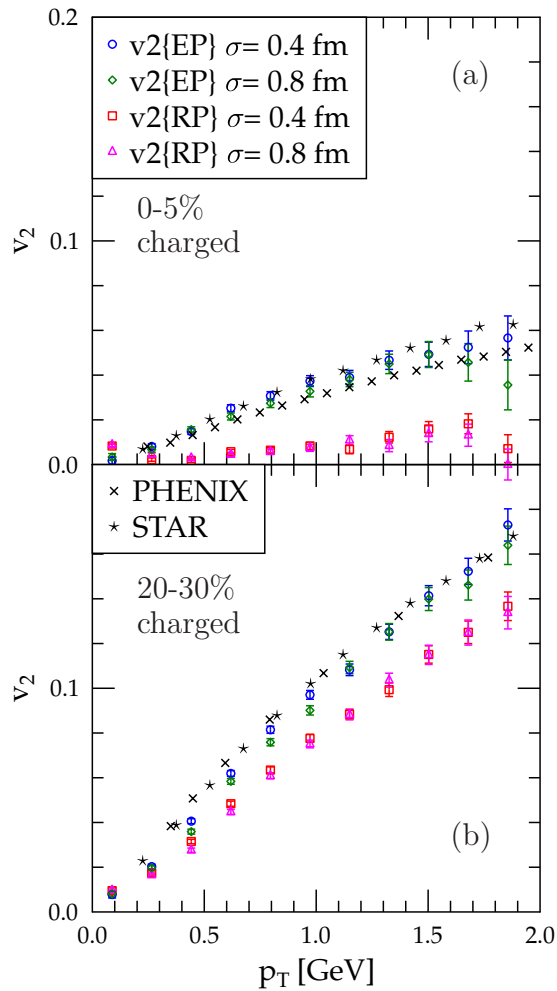}
  \caption{\protect\small (Color online) Elliptic flow of charged particles as a function
           of $p_T$ at different centralities with two different values for Gaussian
           smearing parameter $\sigma$.}
  \label{fig: differential v2 sigma}
\end{figure}

In Fig.~\ref{fig: differential v2} we plot the elliptic flow of charged particles as a
function of $p_T$ at different centralities. We show the event-by-event results for
$v_2\{\text{EP}\}$, $v_2\{\text{RP}\}$ and $v_2\{\text{PP}\}$, as well as
$v_2\{\text{RP,ave}\}$ which is obtained from averaged initial states.

First, we observe, that $v_2\{\text{RP}\}$ and $v_2\{\text{RP,ave}\}$ are quite close
to each other (although in the panel (c) some statistical fluctuations seem to be still
present), and especially that in central collisions there is a significant deficit of
$v_2$ relative to the data. Second, we see that our $v_2\{\text{EP}\}$ agrees very well
(within the estimated errors) with the data up to $p_T\sim2$~GeV in all centrality
classes. Notice also the difference between the uncorrected $v_2\{\text{obs}\}$ and the
corrected, final, $v_2\{\text{EP}\}$; especially for central collisions, the corrections
$\mathcal{R}_2$ are quite large. Thus, fluctuations alone are not sufficient in
explaining the $v_2$ data but that -- in addition to taking into account the
fluctuations -- the computed $v_2$ must be defined in the same way as in the experimental
analysis.  

Third, we notice that the relative increase from $v_2\{\text{RP}\}$ to $v_2\{\text{EP}\}$
decreases from central to peripheral collisions: $v_2\{\text{EP}\}/v_2\{\text{RP}\} =
{\cal O}(10)$ in panel (a) and ${\cal O}(1.2)$ in panel (d). Fourth, contrary to our
original expectation, $v_2\{\text{RP,ave}\}$ for semi-peripheral collisions is actually
below (and not above) the data at $p_T \sim 1.5$~GeV. This is due to the fact that with
our MCG model and smearing, the actual energy density profiles become flatter and less
eccentric than the conventional eWN profiles obtained from an optical Glauber model.
As a result, we get a smaller $v_2\{\text{RP}\}$ than e.g. in Ref.~\cite{Niemi:2008ta},
and thus also in the 20-30\% centrality class there is room  for an increase from
$v_2\{\text{RP}\}$ to $v_2\{\text{EP}\}$. From these observations we can conclude that
we have answered the questions {\em (i)--(iii)} presented in Sec.~\ref{intro}.

Fourth, Fig.~\ref{fig: differential v2} indicates that $v_2\{\text{PP}\}$ is very
close to $v_2\{\text{EP}\}$ in all centrality classes. This result suggests that the
participant plane indeed is quite a good approximation for the event plane.

In Fig.~\ref{fig: differential v2 sigma} we show the effects of varying our Gaussian
smearing parameter $\sigma$. We see that our elliptic flow results are quite insensitive
to $\sigma$: Doubling the value of $\sigma$ causes only of the order of 10\% changes in
our $v_2(p_T)$. We remind, however, that our  $p_T$-spectra and multiplicity of pions
were not as stable against $\sigma$ but we expect that doing more proper fitting to the
pion spectra by fine-tuning $\Tdec$ and the initial overall normalization constant $K$,
would not affect our $v_2$ results significantly. 

\begin{figure}[b!]
  \includegraphics[height=9.0cm]{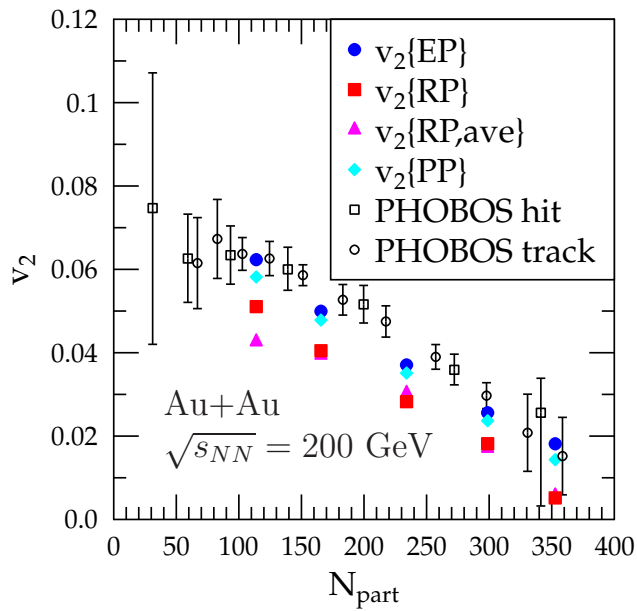}
  \caption{\protect\small (Color online) Integrated elliptic flow for Au+Au
           collisions at $\sqrt{s_{NN}} = 200 \GeV$ calculated with fluctuating and
           averaged initial states are shown for $\sigma=0.4\fm$. Data from the PHOBOS
           collaboration \cite{Alver:2006wh} are shown with statistical and systematic
           errors added in quadrature.}
  \label{fig: int v2}
\end{figure}

In Fig.~\ref{fig: int v2} we plot the integrated elliptic flow for the four different
cases considered above and the data from the PHOBOS collaboration \cite{Alver:2006wh}.
As expected on the basis of Figs.~\ref{fig: differential v2} and \ref{fig: pion spectra},
our results $v_2\{\text{EP}\}$ and $v_2\{\text{PP}\}$ now agree with the data very well,
while the $v_2\{\text{RP}\}$ results fall significantly below the data. 

\begin{figure}[t]
  \includegraphics[height=9.0cm]{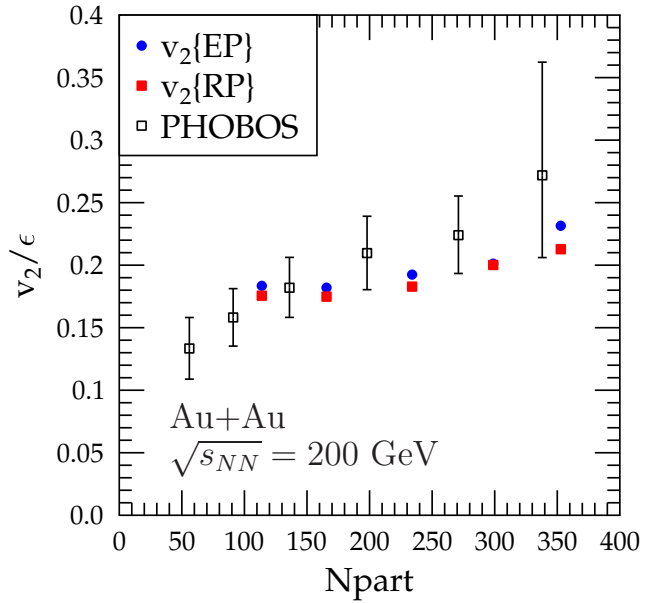}
  \caption{\protect\small (Color online) Integrated elliptic flow of charged particles
           for Au+Au collisions at $\sqrt{s_{NN}} = 200 \GeV$ divided by initial
           eccentricity. Theoretical calculations correspond to fluctuating 
           initial states with $\sigma=0.4\fm$. Data from the PHOBOS collaboration
           \cite{Alver:2006wh} are shown with statistical and systematic errors added
           in quadrature.}
  \label{fig: int v2/ecc}
\end{figure}

Next, Fig.~\ref{fig: int v2/ecc} shows the computational quantity $v_2/\epsilon$ 
which is often discussed. In the PHOBOS result \cite{Alver:2006wh}, $v_2$ is determined
relative to the event plane while the initial state eccentricity is computed relative
to the participant plane. We reproduce the PHOBOS $v_2/\epsilon$ if we do the same,
i.e. use $\epsilon_{\text{PP}}$ from Eq.~\eqref{eq:eccPP}. Interestingly, if we replace
both the elliptic flow and the eccentricity by their reaction plane analogues, we can
still get a scaling law that agrees with our $v_2\{\text{EP}\}/\epsilon_{\text{PP}}$
and with the data. This figure illustrates again the importance of the consistency
in the reference plane definition.

\begin{figure}[h]
  \includegraphics[height=9.0cm]{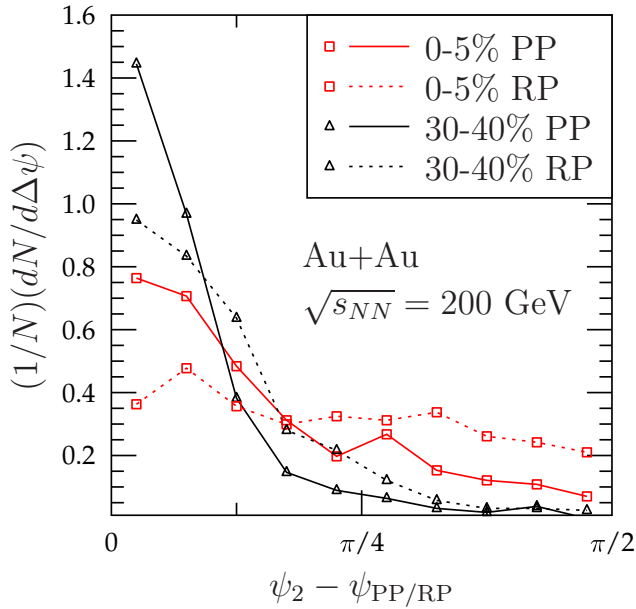}
  \caption{\protect\small (Color online) Correlation of the event plane with the
           participant plane, and with the reaction plane at different centralities and
           with $\sigma=0.4\fm$. The lines are to guide the eye.}
  \label{fig: ep vs rp}
\end{figure}

Finally, we answer the question {\em (iv)} presented in Sec.~\ref{intro}. 
Figure~\ref{fig: ep vs rp} shows the correlation between the event plane and the
participant plane as well as the correlation between the event plane and the reaction
plane. We plot the distribution of events as a function of  the angle differences
$\psi_{2}-\psi_{\text{PP}}$ and $\psi_{2}-\psi_{\text{RP}}$. For this figure, we have
used each hydro run only once. We notice that in central collisions, the planes are
more weakly correlated than in semi-peripheral collisions where clearer peaks around
$\psi_{2}=\psi_{\text{PP}},\psi_{\text{RP}}$ arise. As expected based on
Fig.~\ref{fig: differential v2}, the participant plane is indeed found to be quite
a good approximation for the event plane, in all centrality classes.
However, fluctuations of the event plane around the ''true'' event plane are much
larger in central collisions and thus the correlation between the event plane and
the participant plane in Fig.~\ref{fig: ep vs rp} looks weaker for central collisions
than for the more peripheral ones.

\section{Conclusions}

The main result of this paper is that using event-by-event ideal hydrodynamics with
MCG-generated  fluctuating initial density profiles we can simultaneously reproduce
the measured centrality dependence and the $p_T$ shape of charged-particle elliptic
flow up to $p_T\sim2$~GeV. Also the measured pion spectra are quite well reproduced,
although we have not made an effort to finetune the model parameters. In particular,
in addition to accounting for the fluctuations in the system, we have demonstrated
the importance of using the same $v_2$ definition as in the data analysis.  

We have performed all hydrodynamic simulations with zero viscosity. Thus, our results
suggest that extracting a non-zero lower limit for the viscous coefficients from the
measured $v_2(p_T)$ of charged hadrons is practically impossible without further
constraints to the model, especially to the initial state. We would like to emphasize,
that we have for simplicity considered only the event-by-event analogues of the eWN
initial profiles whose eccentricities are typically smaller than e.g. those of the
eBC- or CGC-type \cite{Lappi:2006xc,Drescher:2006ca,Drescher:2007ax} of profiles.
Whether the data are still consistent with non-zero viscosity with these initial
conditions, is left as a future exploration. Nevertheless, our results demonstrate
that event-by-event hydrodynamics with initial state fluctuations must be accounted
for before a more reliable lower limit for viscosity can be obtained from elliptic
flow data. 

We have shown that the definition of the reference plane with respect to which one
determines  $v_2$, plays an important role especially in central collisions. On the one
hand, if $v_2$ is computed relative to the reaction plane (determined by the impact
parameter), the fluctuating and averaged initial states lead practically to the same
results. In this sense, the previous conventional ideal hydrodynamical results for
the system evolution are still relevant in central enough collisions but one should
not compare the reaction-plane $v_2$ to the event-plane $v_2$ quoted by the experiments.
On the other hand, according to our results, the initial participant plane seems to be
quite a good approximation for the event plane in the presence of hydrodynamically
evolving density fluctuations.

The present work can obviously be improved in many ways. Especially in event-by-event
hydrodynamics the decoupling temperature may vary from event to event. Instead of a
fixed $\Tdec$ applied here, one could implement a dynamical freeze-out criterion as
was done e.g. in Ref.~\cite{Eskola:2007zc}. However, in order to improve upon the
well-known problem of the proton $p_T$ spectra when partial chemical equilibrium is
not applied, one could couple our hydro to a hadron cascade afterburner which would
handle also the resonance decays \cite{Werner:2010aa,Petersen:2009vx,Petersen:2010md}.
Related to the initial state, one should more closely inspect the uncertainties due
the assumed energy density smearing, which is an avoidable issue with event-by-event
hydrodynamics. Here we found out that $v_2$ remained fairly insensitive to Gaussian
smearing width while pion $p_T$ spectra were more sensitive to it towards larger $p_T$.
Also other possible smearing functions should be studied. One should also consider
a dynamical QCD-based model for the initial  fluctuations, in which case also the
absolute initial density profiles should be computable. These tasks, and considering
the effects of fluctuations on other observables, we leave as future developments.

\begin{acknowledgments}
We gratefully acknowledge the financial support from the Academy of Finland, KJE's
projects 115262 and 133005, from the national Graduate School of Particle and Nuclear
Physics (HH), and from the University of Jyv\"askyl\"a (HH's grant). The work of HN was
supported by the Extreme Matter Institute (EMMI). HH thanks the [Department of Energy's]
Institute for Nuclear Theory at the University of Washington for its hospitality and
the Department of Energy for partial support during the completion of this work. We
thank D.H. Rischke, P.V. Ruuskanen, M. Luzum and J.-Y. Ollitrault for helpful
discussions. All the supercomputing was done at the CSC -- IT Center for Science in
Espoo, Finland.
\end{acknowledgments}

\end{document}